\documentclass[11pt]{article}
 
    \usepackage[a4paper]{geometry} 
    \usepackage[utf8]{inputenc}  
    \usepackage{enumerate}
    \usepackage{mathtools}
    \usepackage{braket}
    \usepackage{eufrak}
    \usepackage{commath} 
    \usepackage{amsmath}   
    \usepackage{multicol}
    
    \let\Pi\varPi
    \let\epsilon\varepsilon
    
    \usepackage{subcaption}
    \usepackage{bm}
    \usepackage{bbm}
    \usepackage{amssymb} 
    \usepackage{mathrsfs}
    \usepackage{graphicx}
    \usepackage{appendix}
    \usepackage{csquotes} 
    \usepackage{parskip}
    \usepackage{hyperref}
    \hypersetup{
        colorlinks   = true,        
        urlcolor     = blue,        
        linkcolor    = black,       
        citecolor   = blue          
    }
    \usepackage[numbers,sort&compress]{natbib} 

    \DeclareMathAlphabet{\mathpzc}{OT1}{pzc}{m}{it} 

    \newcommand{\mr}[1]{\mathrm{#1}}

    \newcommand{\mbf}[1]{\mathbf{#1}}   
    
    \newcommand{\bk}{\mathbf{k}}
    \newcommand{\ehat}{\bm{\hat{e}}}

\title{The Faraday Effect as a Two-State Quantum Process}   

\author{
    Benjamin W. Butler\\[1.0ex]
    Quantum Theory Group, University of Glasgow, United Kingdom\\[1.0ex]
    \texttt{b.butler.1@research.gla.ac.uk}
}

\date{}

\begin{document}

\maketitle

\begin{abstract}
    We show that the Faraday effect can be derived from a simple two-state model. This approach uses a quantized electromagnetic field and does not make reference to differences in refractive indices of left- and right-circularly polarized light. Instead it treats the effect as a forward Rayleigh scattering process between two orthogonal modes of the quantized field, and thus emphasises the quantum-mechanical aspects of the phenomenon. 
\end{abstract}

\begin{multicols}{2}

When linearly polarized light passes through a medium in the presence of a parallel static magnetic field it experiences a rotation of its polarization direction. This is known as magnetic optical rotation (OR) or the Faraday effect \cite{faraday_experimental_1846,barron_molecular_2004}. OR can also occur in the absence of a magnetic field, for example if the medium is rotating \cite{jones_rotary_1976}, or if it is composed of chiral molecules. The latter is known as natural OR \cite{barron_molecular_2004}. Power and Thirunamachandran have previously shown that it is possible to derive the angle of OR due to chiral molecules using a simple two-state model \cite{power_optical_1971}. Here we extend this approach to the Faraday effect. 

Classically the Faraday effect is understood to arise as a result of circular birefringence \cite{barron_molecular_2004}, i.e., a difference in refractive index for left- and right-circularly polarized light due to the presence of the magnetic field. A modern approach to calculating the angle of rotation can be found in the book by Barron \cite{barron_molecular_2004}. There the molecules are treated quantum mechanically, the light classically, and the interaction developed through molecular property tensors. We follow Power and Thirunamachandran's lead and take a completely different approach, which does not make reference to birefringence explicitly, and treats the light quantum mechanically.

In particular, we treat the electromagnetic (EM) fields associated with the light as quantum field operators, whilst still treating the molecules using the usual Schrödinger theory. This is a regime known as molecular or non-relativistic quantum electrodynamics (QED) \cite{craig_molecular_1998,cohen-tannoudji_photons_1989}. In this regime the Hamiltonian of the total system, i.e., molecules plus radiation field, is  
\begin{equation}
    \hat{H} = \sum_{\zeta = 1}^N \hat{H}_{\mathrm{mol}}(\zeta) + \sum_{\bk \lambda} \hat{H}_{\mr{rad}}(\bk, \lambda) + \hat{H}_{\mathrm{int}}.
\end{equation}
Here $\hat{H}_{\mathrm{mol}}(\zeta)$ is the (non-relativistic) Hamiltonian of the $\zeta$th molecule, taken in the Born-Oppenheimer approximation \cite{bunker_fundamentals_2005}, and $\hat{H}_{\mathrm{rad}}(\bk, \lambda)$ is the Hamiltonian of the $(\bk, \lambda)$ mode of the EM field, where $\bk$ denotes the wave vector and $\lambda$ the polarization of the mode. The eigenstates of $\hat{H}_{\mathrm{rad}}(\bk, \lambda)$ are the \enquote{number states} $\ket{n(\bk,  \lambda)}$, where $n$ indicates the occupation number of the mode. These two Hamiltonians act wholly on separate state spaces and are in this sense decoupled from one another. The interaction Hamiltonian, $\hat{H}_{\mathrm{int}}$, is needed to couple the two systems. The interaction can take different forms depending upon the situation under consideration. If the wavelength of the radiation is large compared to the dimensions of the molecules then it is sufficient in many cases to work in the \enquote{electric dipole approximation} \cite{craig_molecular_1998}, in which 
\begin{equation} \label{ED int Ham}
    \hat{H}_{\mr{int}} = - \sum_{\zeta} \hat{\bm{\mu}}(\zeta) \cdot \hat{\mathbf{E}}(\mathbf{R}_{\zeta}), 
\end{equation}
where $\hat{\bm{\mu}}(\zeta)$ is the electric dipole operator of the $\zeta$th molecule and $\hat{\mathbf{E}}(\mathbf{R}_{\zeta})$ the electric field \emph{operator} evaluated at this molecule's centre of mass. The electric field operator is \cite{craig_molecular_1998}
\begin{equation}
\hat{\mbf{E}}(\mbf{r}) = i \sum_{\mbf{k}, \lambda} \left( \frac{\hbar c |\bk|}{2 \varepsilon_0 V} \right)^{1/2} \left[ \ehat_{\mbf{k} \lambda} e^{i \mbf{k} \cdot \mbf{r}} \hat{a}_{\mbf{k} \lambda} - \ehat^*_{\mbf{k} \lambda} e^{- i \mbf{k} \cdot \mbf{r}} \hat{a}^{\dagger}_{\mbf{k} \lambda}  \right],
\end{equation}
where $\ehat_{\mbf{k} \lambda}$ is the unit polarization vector associated with mode $(\bk, \lambda)$, and $\hat{a}_{\mbf{k} \lambda}$, $\hat{a}_{\mbf{k} \lambda}^{\dagger}$ are the annihilation and creation operators, respectively, for this same mode. The actions of these operators on the number states are 
\begin{subequations}
    \begin{gather}
        \hat{a}_{\mbf{k} \lambda} \ket{n(\bk, \lambda)} = \sqrt{n} \ket{(n-1)(\bk, \lambda)}, \\
        \hat{a}_{\mbf{k} \lambda}^{\dagger} \ket{n(\bk, \lambda)} = \sqrt{n+1} \ket{(n+1)(\bk, \lambda)}.
    \end{gather}
\end{subequations}
Clearly $\hat{a}_{\mbf{k} \lambda}$ is physically associated with an absorption and $\hat{a}_{\mbf{k} \lambda}^{\dagger}$ an emission of a photon of mode $(\bk, \lambda)$. Note that the creation and annihilation operators of one mode do not act on the states of any other mode.

In an OR experiment light is elastically scattered from one mode, henceforth known as mode 1, into another mode, mode 2. The two modes have the same wave vector ($\bk$) but orthogonal linear polarizations ($\ehat^{(1)}$ and $\ehat^{(2)}$, say). The resulting superposition between the two modes makes it appear as though the light's polarization has rotated through an angle $\theta$. Classically this angle can be defined via  
\begin{equation}
    \tan \theta = \sqrt{\frac{I_2 }{ I_1}}
\end{equation}
where $I_1$ is the intensity of light scattered into mode 1 and $I_2$ is the intensity scattered into mode 2. In a quantum mechanical-picture we can replace the intensities with the expectation values of the number operator for each mode, so that 
\begin{equation} \label{Quantum angle defn}
    \tan \theta = \sqrt{\frac{\braket{\hat{n}_2}}{ \braket{\hat{n}_1}}}.
\end{equation}
A quantum derivation of OR (magnetic or natural) therefore rests on finding these expectation values. 

Power and Thirunamachandran \cite{power_optical_1971} began by considering the natural OR due to a single molecule. After the scattering there is a non-zero probability of mode 2 being occupied by any number of photons from zero to $n$. However, it is most probable that either no photon is transferred to mode 2, or a single photon is. Non-forward scattering is also possible but is suppressed as compared to the forward process. Therefore Power and Thirunamachandran make a two-state approximation, akin to the one commonly found in quantum optics \cite{allen_optical_1987,frasca_modern_2003}, and assume that the dynamics of the combined system are dominated by the states  
\begin{subequations}
    \begin{gather}
        \ket{a} \equiv \ket{n(1)}, \\
        \ket{b} \equiv \ket{(n-1)(1); 1 (2)},
    \end{gather}
\end{subequations}
where the molecular state has been suppressed as it is the same in $\ket{a}$ and $\ket{b}$, and a simplified notation has been introduced for the number states. The state of the system at some time $t$ can be written as 
\begin{equation}
    \ket{\Psi(t)} = c_a(t) \ket{a} + c_b(t) \ket{b},
\end{equation}
where $c_{a/b}(t)$ are the probability amplitudes for being in either state. The probabilities can be found analytically by solving 
\begin{equation} \label{Effective Sch eqn}
    i \hbar \frac{\partial}{\partial t} \ket{\Psi(t)} = \hat{H}_{\mr{eff}} \ket{\Psi(t)},
\end{equation}
where 
\begin{equation} \label{two state Ham}
    \hat{H}_{\mr{eff}} = \begin{pmatrix}
        E_0 & M^* \\ M & E_0
    \end{pmatrix}
\end{equation}
is the effective Hamiltonian for the two-state system. $M$ is the matrix element between $\ket{a}$ and $\ket{b}$ and $E_0$ is the energy of both states. Assuming the initial state is $\ket{a}$ we find 
\begin{subequations}
    \begin{gather}
        | c_a(t) |^2 = \cos^2 \left( |M| t / \hbar \right), \\
        | c_b(t) |^2 = \sin^2  \left( |M| t / \hbar \right).
    \end{gather}
\end{subequations}
Using this result together with (\ref{Quantum angle defn}) we find that, in the small angle approximation, the angle of rotation is given by    
\begin{equation} \label{single molecule angle}
    \theta \approx \frac{|M| t }{ \hbar \sqrt{n}}.
\end{equation}

If instead of a single molecule we consider a gas then this result requires slight modification. If we assume the gas is dilute and the number of molecules interacting with the optical field is $N$ then, because forward scattering is a coherent process \cite{craig_molecular_1998}, the relevant matrix element is $N M$, $M$ being the transition amplitude in the single-molecule case. If the length of the sample is $L = ct$ then the total optical rotation angle due to the gas may be written as 
\begin{equation} \label{angle due to gas}
    \theta \approx \frac{N |M| L}{ \hbar c \sqrt{n} }.
\end{equation}
This is the central result of Power and Thirunamachandran's work \cite{power_optical_1971}. The calculation of the OR angle has been reduced to the calculation of the quantum transition amplitude $M$ from state $\ket{a}$ to $\ket{b}$.

In natural OR the electric dipole interaction Hamiltonian (\ref{ED int Ham}) is not sufficient, as the effect occurs through the mixing of electric and magnetic dipole terms in the matrix element $M$. Therefore Power and Thirunamachandran use the (single-molecule) interaction Hamiltonian \cite{power_optical_1971}
\begin{equation} \label{NOR int ham}
    \hat{H}_{\mr{int}}^{\mr{NOR}} = - \hat{\bm{\mu}} \cdot \hat{\mbf{E}} - \hat{\mbf{m}} \cdot \hat{\mbf{B}},
\end{equation}  
where $\hat{\mbf{m}}$ is the magnetic dipole moment and $\hat{\mbf{B}}$ the magnetic field operator (associated with the incident light field). They then calculate $M$ using standard second order perturbation theory \cite{craig_molecular_1998,messiah_quntum_1961},
\begin{equation} \label{2nd order matrix element}
    M = \sum_{I} \frac{\braket{b | \hat{H}_{\mr{int}} | I} \braket{I | \hat{H}_{\mr{int}} | a}}{E_a - E_I},
\end{equation}
where the summation is over all possible intermediate states of the combined system and the denominator contains the energies of the states. 

In the Faraday effect we have a static magnetic field $\mbf{B}$ present, as well as the incident light field. External fields such as $\mbf{B}$ are not associated with the radiation field and are thus not quantized \cite{craig_molecular_1998}. Therefore the relevant interaction Hamiltonian is 
\begin{equation} \label{FE int ham}
    \hat{H}_{\mr{int}}^{\mr{FE}} = - \hat{\bm{\mu}} \cdot \hat{\mbf{E}} - \hat{\mbf{m}} \cdot \mbf{B},
\end{equation}
where the quantized magnetic field due to the radiation is now ignored as it plays no part in the Faraday effect. It is again the mixing between electric and magnetic terms which gives rise to the effect. Given that (\ref{FE int ham}) \emph{looks} almost identical to (\ref{NOR int ham}) it would be reasonable, perhaps, to assume that we can also just plug this into the perturbation expansion and find the matrix element. However, this is not the case.  The problem is that the static magnetic field cannot create or destroy photons and therefore cannot connect the states $\ket{a}$ and $\ket{b}$ to second order (recall that a transition from $\ket{a}$ to $\ket{b}$ involves the absorption of one photon from mode 1 and the emission of a photon into mode 2). 

To connect the states using a mixture of electric and (static) magnetic terms we require at least two intermediate states, which corresponds to the third order expression for the matrix element \cite{craig_molecular_1998},
\begin{equation}
    M = \sum_{I_1} \sum_{ I_2} \frac{\braket{b |\hat{H}_{\mr{int}} | I_2 } \braket{I_2 | \hat{H}_{\mr{int}} | I_1} \braket{I_1 | \hat{H}_{\mr{int}} | a}}{(E_{I_1} - E_a)(E_{I_2} - E_a)}.
\end{equation}
The sums are again over the possible intermediate states $I_1$ and $I_2$. This expression is more promising but again suffers from a fatal flaw: it cannot deal with degenerate states. If the atom/molecule contains any degeneracy then the denominator will become zero and therefore this formula is not applicable. For the Faraday rotation to be calculated via (\ref{angle due to gas}) a new procedure is required, which we outline below.

\end{multicols}

\begin{figure}[t!] 
    \centering
    \begin{subfigure}[t]{0.33\textwidth} 
        \centering
        \includegraphics[width=\textwidth]{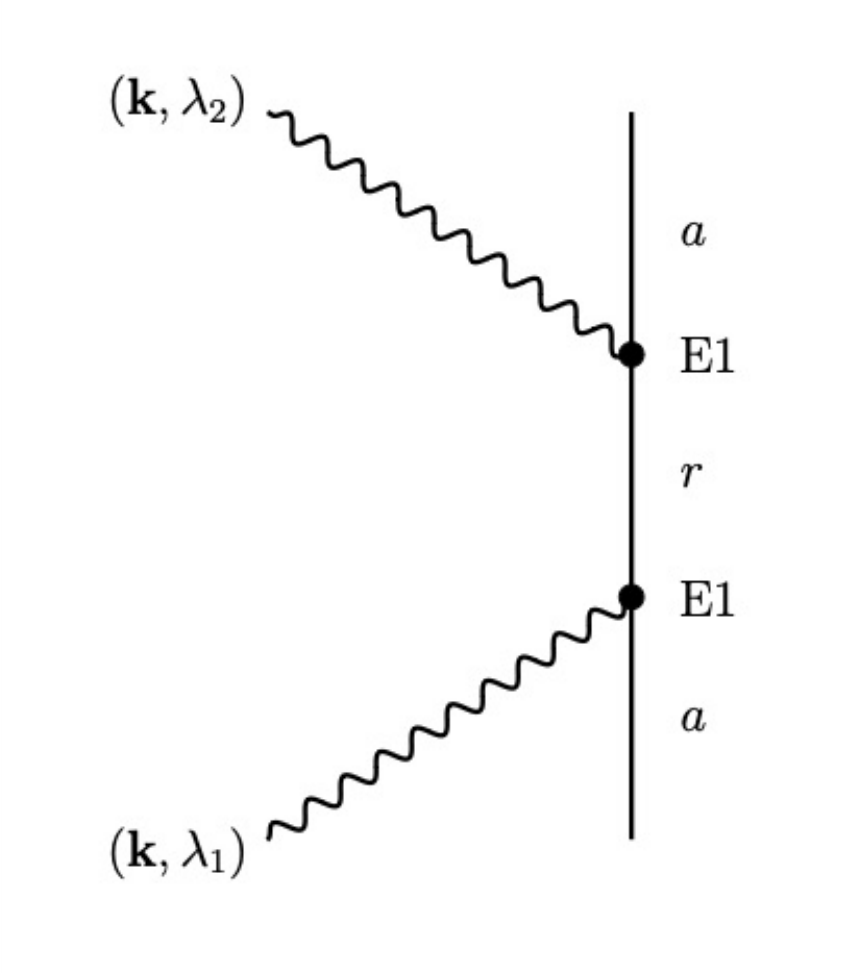}
        \caption{}
        \label{Type a}
    \end{subfigure}%
    ~
    \begin{subfigure}[t]{0.33\textwidth} 
        \centering
        \includegraphics[width=\textwidth]{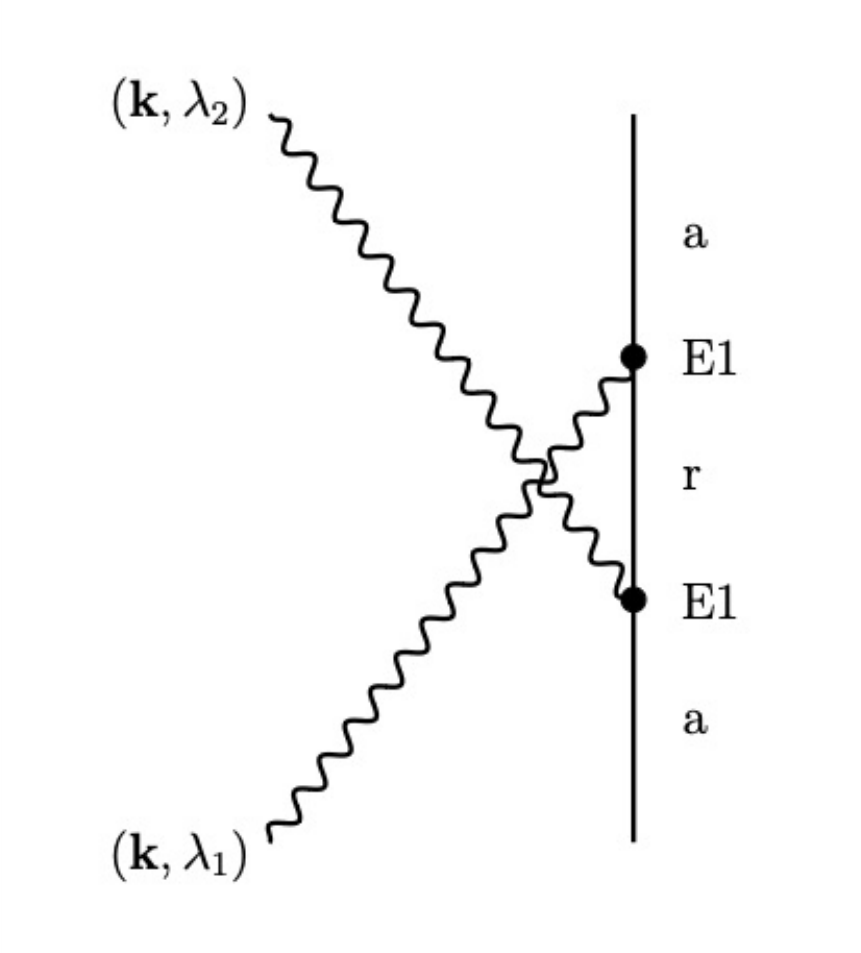}
        \caption{}
        \label{type b}  
    \end{subfigure}%
    \caption{Time-ordered graphs (or Feynman diagrams) showing the interaction between the molecule and the light. Time proceeds vertically in the diagram. The solid vertical line indicates the state of the molecule at a particular time. The emission or absorption of a photon is indicated by a wavy line and is accompanied by a corresponding change in the state of the molecule via the electric dipole (E1) interaction. In diagram (a) the absorption of a photon from mode 1 precedes the emission of a photon into mode 2. In diagram (b) the opposite time ordering is shown.}
    \label{Feynman diagrams}
\end{figure}

\begin{multicols}{2}
We begin by ignoring the light field. The Hamiltonian is then (for the single molecule)
\begin{equation}
    \hat{H}_{\mr{mol}}' = \hat{H}_{\mr{mol}} - \hat{\mbf{m}} \cdot \mbf{B}.
\end{equation}
Let's assume that the eigenbasis of $\hat{H}_{\mr{mol}}$ is known and label the eigenstates and energies by $\ket{E_i^{(0)}}$ and $E_i^{(0)}$, respectively. Let's also assume that the static magnetic field is sufficiently weak such that it can be treated as a perturbation on the molecule. Therefore we expect the molecule's state and energy in the presence of the perturbation to be modified by a small amount, which can be encapsulated in a power series expansion in terms of some order parameter, $\delta$ say. To lowest order the perturbed states and energies are 
\begin{subequations} \label{State and energy to first order}
    \begin{gather}
        \ket{E_i} = \ket{E_i^{(0)}} + \delta \ket{E_i^{(1)}}, \\
        E_{i} = E_{i}^{(0)} + \delta E_i^{(1)}.
    \end{gather}
\end{subequations}
As is well known, the calculation of the first order corrections to the states and energies, i.e., $\ket{E_i^{(1)}}$ and $E_i^{(1)}$, depends on whether the initial state is degenerate or not. Assuming that the molecule in question does not contain any degeneracy then we have \cite{messiah_quntum_1961,atkins_molecular_2010}
\begin{subequations} \label{state and energy corrections}
    \begin{gather}
        \ket{E_i^{(1)}} = \sum_{n \neq i} \ket{E_n^{(0)}} \frac{\braket{E_n^{(0)} | V | E_i^{0}}}{E_i^{(0)} - E_n^{(0)}}, \\
        E_i^{(1)} = \braket{E_i^{(0)} | V | E_i^{(0)}},
    \end{gather}
\end{subequations}
where, in our case, $V = - \hat{\mbf{m}} \cdot \mbf{B}$. 

Let's reintroduce the optical field. The Hamiltonian of the combined system, in the QED formalism, may now be written as 
\begin{eqnarray*}
    \hat{H} = \hat{H}_{\mr{mol}}' + \hat{H}_{\mr{rad}} + \hat{H}_{\mr{int}}.
\end{eqnarray*}
The eigenstates of $\hat{H}_{\mr{mol}}' + \hat{H}_{\mr{rad}}$ are product states of $\ket{E_i}$ and the number states, where to first order $\ket{E_i}$ are given by (\ref{State and energy to first order}) and (\ref{state and energy corrections}). 

We will find it sufficient now to use the electric dipole form of the interaction Hamiltonian (\ref{ED int Ham}) and the second order expansion for $M$ (\ref{2nd order matrix element}). The possible intermediate states are $\ket{I} = \ket{E_r ; (n-1)(1)}$ and $\ket{I} = \ket{E_r ; n (1) ; 1(2)}$, where $\ket{E_r}$ is a state of the (perturbed) molecule. The calculation of $M$ using (\ref{2nd order matrix element}) is facilitated by the use of the time-ordered graphs shown in Fig \ref{Feynman diagrams}. After some simple rearrangement we find 
\begin{multline} \label{M rearranged}
    M = \frac{\mu_0 c^2 \hbar \omega \sqrt{n}}{V} \hat{e}_i^{(1)} \hat{e}_j^{(2)} \\
    \times \sum_{r} \frac{E_{rg} \mr{Re}\{ \mu_i^{gr} \mu_j^{rg} \} - i \hbar \omega \mr{Im} \{ \mu_i^{gr} \mu_j^{rg} \} }{(\hbar \omega)^2 - E_{rg}^2}.
\end{multline}
In this expression we have the following quantities: the frequency of the optical radiation, $\omega = c | \mbf{k} |$, the vacuum permeability $\mu_0$, the energy difference $E_{rg} = E_r - E_g$, where $E_g$ is the (perturbed) ground state energy of the molecule, and the components of the electric dipole transition moments between the states, $\mu_i^{gr} \equiv \braket{E_g | \hat{\mu}_i | E_r}$ etc. The repeated indices ($i$ and $j$) are summed over implicitly, and the sum over $r$ is over the perturbed states of the molecule. To bring the magnetic field into this expression we need to replace the states $\ket{E_i}$ with the approximate forms given by (\ref{State and energy to first order}). For example, to first order in $\delta$ the dipole transition moments become 
\begin{multline} \label{expanded transition moment}
    \bm{\mu}^{mn} = \braket{E_m^{(0)} | \hat{\bm{\mu}} | E_n^{(0)}} \\+ \delta \left( \braket{E_m^{(1)} | \hat{\bm{\mu}} | E_n^{(0)}} + \braket{E_m^{(0)} | \hat{\bm{\mu}} | E_n^{(1)}} \right).
\end{multline}

The exact result will depend on whether we assume the molecule is degenerate or not. As is well known the Faraday rotation can be expressed as the sum of three terms \cite{serber_theory_1932,barron_molecular_2004}, the so-called Faraday \enquote{A}, \enquote{B}, and \enquote{C} terms. The A and C terms are only present if the molecule contains degeneracy \cite{barron_molecular_2004}. For simplicity we will assume a non-degenerate molecule and therefore only be concerned with deriving the B term. Under this assumption we can use (\ref{state and energy corrections}) for the corrections to the states and energies.

After some simplification of (\ref{M rearranged}) using (\ref{expanded transition moment}) and (\ref{state and energy corrections}) we find that the angle of rotation (\ref{angle due to gas}) can be written as  
\begin{multline}
    \theta \approx - \frac{\mu_0 c L \eta (\mbf{B} \cdot \mbf{k})}{\hbar | \mbf{k} |}  \sum_r \left[ \left( \frac{\omega^2}{\omega_{rg}^{(0) \, 2} - \omega^2} \right) \right. \\
    \mr{Im} \left\{ \sum_{p \neq g} \frac{m_3^{pg(0)}}{\hbar \omega_{pg}^{(0)}} ( \mu_1^{gr(0)} \mu_2^{rp(0)} - \mu_2^{gr(0)} \mu_1^{rp(0)}  ) \right. \\
    \left. \left. ~ ~ ~ + \sum_{s \neq r} \frac{m_3^{rs(0)}}{\hbar \omega_{sr}^{(0)}} ( \mu_1^{gr(0)} \mu_2^{sg(0)} - \mu_2^{gr(0)} \mu_1^{sg(0)} ) \right\} \right],
\end{multline}
where $\eta = N/V$ is the number density of molecules in the gas, $\omega_{mn}^{(0)} = (E_m^{(0)} - E_n^{(0)}) / \hbar$, the transition moments $\mu_i^{mn(0)}$, $m_i^{mn(0)}$ are taken with respect to the unperturbed molecular states, and the subscripts $1$, $2$, $3$ refer to components in the direction of $\ehat^{(1)}$, $\ehat^{(2)}$ and $\mbf{k}$, respectively. This is the form of the Faraday B term found in the literature \cite{barron_molecular_2004,serber_theory_1932}, thus showing the equivalence of our approach to previous ones. Again we wish to emphasise that our method can easily be extended to the case of degenerate states.

In summary, we have extended a method due to Power and Thirunamachandran \cite{power_optical_1971} for calculating natural OR angles by incorporating external perturbations into their theory. By this method we have shown that the Faraday effect can be described as a quantum transition between two orthogonal states of the quantized EM field, with virtual transitions induced by the external magnetic field. We believe this method will serve as a useful alternative to other approaches, such as that found in Barron \cite{barron_molecular_2004}, because it may be more easily accessible to researchers and students more familiar with fundamental quantum optics than methods based on molecular property tensors, and because it highlights the quantum features of an effect which is often considered at the classical level. Application to the mechanical Faraday effect is an avenue we are actively exploring.

I would like to acknowledge financial support from EPSRC grant EP/T517896/1 and a Tandem Fellowship from the European Centre for Advanced Studies.

\bibliography{refs.bib}
\bibliographystyle{ieeetr} 

\end{multicols}

\end{document}